\documentclass[pra,a4,twocolumn,showpacs,amsmath,amssymb]{revtex4-1}

\usepackage{graphicx,subfigure,epsfig}
\usepackage{amssymb}
\usepackage{amsthm}
\usepackage{color}
\usepackage{epstopdf}
\usepackage{epsfig}

\begin{document}

\title{Microgap thermophotovoltaic systems with low emission temperature and high electric output}

\author{M.~S.~Mirmoosa}
\email{mohammad.mirmoosa@aalto.fi}
\affiliation{%
	Department of Radio Science and Engineering, School of Electrical Engineering, Aalto University, P.~O.~Box 13000, FI-00076 AALTO, Finland
}%

\author{M.~Omelyanovich}
\affiliation{%
	Department of Radio Science and Engineering, School of Electrical Engineering, Aalto University, P.~O.~Box 13000, FI-00076 AALTO, Finland
}%
\affiliation{%
	Department of Nanophotonics and Metamaterials, ITMO University, 197043 St. Petersburg, Russia
}%

\author{C.~R.~Simovski}
\affiliation{%
	Department of Radio Science and Engineering, School of Electrical Engineering, Aalto University, P.~O.~Box 13000, FI-00076 AALTO, Finland
}%
\affiliation{%
	Department of Nanophotonics and Metamaterials, ITMO University, 197043 St. Petersburg, Russia
}%

\date{\today}

\begin{abstract}
 We theoretically show that a thermophotovoltaic (TPV) system enhanced by a wire metamaterial opens the door to a prospective microgap thermophotovoltaics which will combine high  electric output with relatively low temperatures of the emitter. The suggested system comprises an array of parallel metal nanowires grown on top of a photovoltaic semiconductor and standing free in the vacuum gap between the host dielectric layer and the emitter, so that their ends are sufficiently close to the emitting surface. Due to the resonant near-field coupling between this wire medium and the emitter and due to the optimized layered structure of the whole system, the strongly super-Planckian radiative heat flux of resonant nature is engineered. 
 \end{abstract}

\maketitle

\section{Introduction}

In thermophotovoltaic (TPV) systems (see e.g.~Refs.~\cite{Bauer,Lenert}), a hot emitter of thermal radiation whose temperature is determined by a heat source (e.g.~flame) transmits infrared photons to the photovoltaic (PV) cell as it is shown in Fig.~1. The cell is formed by a doped semiconductor possessing the significant PV spectral response at infrared frequencies. A free-space gap separates the emitter from the PV cell, preventing the conductive thermal flux which would obviously produce harmful phonons in the PV material. Thermal phonons would suppress the photovoltaic generation \cite{Bauer}, and very high temperature of the PV cell (close to that of the emitter) would correspond to large diode current $J_{\rm diode}$. Both these factors would strongly decrease  the total current $J$ in the load (see the equivalent scheme in Fig.~1), and both of them are prevented by the gap. Most popular TPV systems are so-called far-field ones, in which the gap thickness $d$ is larger than the free-space wavelengths $\lambda_{\rm{L}}$ (the largest one in the model spectrum of the emitted radiation). In these systems the process of radiative heat transfer splits onto two independent stages.

\begin{figure}[t!]
\includegraphics[width=9.0cm]{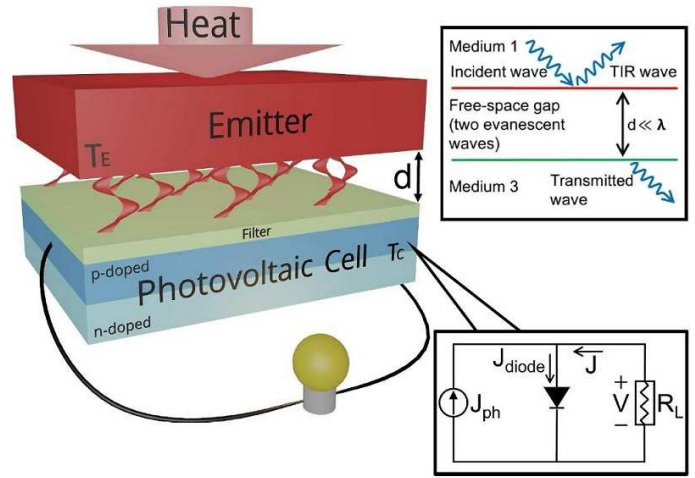}
\caption{Schematic of a far-field TPV system. On the insets: equivalent circuit of the PV cell operating as an electric generator, schematics of the photon tunneling.}
\label{fig:TPV}
\end{figure}

The first stage is thermal radiation governed by the Stefan-Boltzmann and Wien laws. The first law -- ${\cal P}=e\sigma T_{\rm{E}}^4$, where ${\cal P}$ is radiated power per unit area of the emitter, $e$ is its integral emissivity/absorptivity, $\sigma$ is a universal constant ($5.67\times10^{-8}$ in SI units), and $T_{\rm{E}}$ is the temperature of the emitter, -- says that for noticeable electric output the temperature of the emitter must be high enough.  Very high $T_{\rm{E}}$ means very high radiation and promises high electric output. The second law -- $\lambda_{\rm{M}}=b/T_{\rm{E}}$, where $b$ is a universal constant ($2.89\times10^{-3}$ in SI units) delivers the wavelength $\lambda_{\rm{M}}$ at which the radiation spectrum $d{\cal P}/d\lambda$ is maximal (notice that the frequency $\omega_{\rm{M}}$, at which the radiation spectrum $d{\cal P}/d\omega$ attains maximum, is red-shifted compared to $2\pi c/\lambda_{\rm{M}}$). This law poses an additional limitation to the minimally allowed temperatures of the emitter in TPV electric generators. If $T_{\rm{E}}$ is low, the semiconductor of the PV cell must possess a very narrow bandgap $E_{\rm{g}}=\hbar \omega_{\rm{g}}$ ($\omega_{\rm{g}}$ is called the bandgap frequency and $\hbar$ is the Planck constant). Really, if $\omega_{\rm{M}}\ll \omega_{\rm{g}}$, the most part of radiation corresponds to frequencies $\omega<\omega_{\rm{g}}$ and cannot be converted to the photocurrent. Then the PV generation in the structure is low and dissipation not admissible. Practically, one needs $\omega_{\rm{M}}\approx \omega_{\rm{g}}$ \cite{Bauer}. For temperatures $T_{\rm{E}}<400^{\circ}{\rm{K}}$, $\omega_{\rm{M}}=2\pi T_{\rm E}\cdot (5.87\times10^{10})<1.47\times10^{14}$ rad/s, that requires $\omega_{\rm{g}}<1.5\times10^{14}$. Photovoltaic materials with a so narrow bandgap have been recently synthesized, but they cannot operate at room temperatures \cite{Bauer} and require the energy-consuming cooling. As a result, far-field TPV systems operating with such $T_{\rm E}$ cannot serve electric generators. A far-field TPV system whose PV cell operates at room temperatures require $T_{\rm{E}}>600^{\circ}{\rm{K}}$. However, even the temperature $T_{\rm{E}}=600^{\circ}{\rm{K}}$ is not practically sufficient, because the thermal emission governed by the Stefan-Boltzmann law is low at this temperature, and the electric output is poor. Existing far-field TPV systems operate as electric generators at $T_{\rm{E}}\ge 800^{\circ}{\rm{K}}$ \cite{Bauer}. To extract the electric energy from bodies with lower temperatures one has to apply thermoelectric converters whose heat-to-electricity conversion efficiency, in practice, cannot exceed 3\% \cite{Cronin}. Notice, that the PV conversion efficiency in realistic TPV generators operating at $T_{\rm{E}}\approx 2000^{\circ}{\rm{K}}$ attains 20\% \cite{TPV}.

The second stage of the radiative heat transfer process in a far-field TPV system is absorption of radiated power by a PV cell. If the emitter mimics a black-body, its emission is very broadband, and an optical filter with $\omega_{\rm min}=\omega_{\rm{g}}$ shown in Fig.~1 is obviously needed. The filter suppresses the reflection in the operation band of the PV cell (operating as a blooming covering) and fully reflects the emitted harmonics beyond it. In fact, this filter is a bandpass one, moreover it has a relatively narrow pass-band. This is so because the so-called ultimate efficiency of the PV cell decreases versus the band of incident radiation that is called the Shokley-Queisser limit. Low ultimate efficiency implies high dissipation, strong heating, and finally, negligible electric output \cite{Bauer}. Therefore, the most part of the emitted power should be reflected. Often, it is claimed that this energy is not wasted since serves for maintaining the high $T_{\rm{E}}$ in the steady regime. However, the structure is not infinitely extended, as in our idealized sketch (Fig.~1). In reality, there are walls supporting the TPV cavity and there are some losses in the  optical filter. The strong reflection inevitably results in high losses. Therefore, the black-body emitter, though emits the maximal possible power into far zone, is not optimal. For maximal electric output one needs to engineer a narrow-band thermal radiation. This is possible with a resonant  emitter -- nano-textured. nano-patterned or performed as an array of IR nanoantennas \cite{Bauer,Lenert,Shvets}. Its emissivity attains unity at a resonant frequency and is negligibly small beyond a narrow frequency band which is equal to the operation band of an optimized IR PV cell. The efficiency of the PV conversion in such systems is not restricted by the Shokley-Queisser limit and may theoretically achieve 80-90\% (see e.g.~Ref.~\cite{Bauer}).
However, the electric output is determined not only by the efficiency of the conversion, but also by the emitter power. Far-field emission of a resonant emitter keeps fundamentally restricted by the black-body radiation. To maximize the emitted power in the operation band one has to engineer the resonance frequency $\omega_{\rm{res}}$ nearly equal to $\omega_{\rm{M}}$ of the black body having the same temperature. Therefore, far-field TPV systems with resonant emitters are justified only with $T_{\rm{E}}\ge 1500-2000^{\circ}{\rm{K}}$ \cite{Bauer,Lenert,Shvets}.

\section{Problem formulation}

The target of the present work is high electric output from an advanced TPV system whose emitter has relatively low temperature. This is a quite ambitious goal because its practical implementation may result in a technical breakthrough in the field of waste heat harvesting. As a reference value we admit the temperature $T_{\rm{E}}=500^{\circ}{\rm{C}}=773^{\circ}{\rm{K}}$, which is typical for the shell of a car exhausting pipe in the peak regime \cite{TE}. Specially designed thermoelectric generators may convert this waste heat into electricity \cite{TE}.
At this temperature, the output power of the best commercial thermoelectric generator is equal 2.2 W harvested from the area 0.4 m$^2$. This is the value we want to exceed with an advanced TPV system. In other words, we want to design a TPV system with $T_{\rm{E}}=500^{\circ}$C with electric power output noticeably higher than 5.5 W/m$^2$. Since at this temperature $\omega_{\rm M}=2.75\times10^{14}$ rad/s, the most suitable PV material is indium antimonide (InSb) with $\omega_{\rm{g}}=2.58\times10^{14}$ rad/s. It has sufficiently high PV spectral response $R$ at room temperatures, which at frequencies $\omega>\omega_{\rm{g}}$ can be modelled as $R=e_0/\hbar\omega$, where $e_0$ is the electron charge \cite{Bauer}. The operation band of such a PV cell can be nearly $(2.6-4)\times10^{14}$ rad/s that allows the ultimate efficiency higher than $70\%$. However, a far-field TPV system with such a PV cell will not allow the power output higher than 5.5 W/m$^2$ due to two factors: $T_{\rm{E}}=500^{\circ}$C corresponds to low power radiated by the black body in the spectral range $(2.6-4)\times10^{14}$. Black-body power flux integrated over this range is nearly equal 20 W/m$^2$, and this is the maximal power for a resonant emitter. The realistic overall efficiency of the PV cell is below 20\%, that means the output power below 4 W/m$^2$.

The way on which we theoretically attain our goal is the involvement of the near-field RHT. The near-field coupling, also called photon tunneling, is illustrated in the inset of Fig.~1. If the thickness of the free-space (medium 2) gap between hot medium 1 and cold medium 3 (both these media are semitransparent) is subwavelength, the total internal reflection (TIR) is frustrated. A part of the incident wave energy is transferred from the hot medium to the cold one by a couple of mutually coherent evanescent waves. In the black-body limit RHT is restricted by the waves incident at the interface between media 1 and 2 with the angles $\theta<\theta_{{\rm{TIR}}}$. Beyond the black-body limit (if $d\ll \lambda$) this spatial spectrum broadens up to $\theta<\pi/2$ and even more, because waves which are evanescent in medium 1 are also involved. Such RHT becomes super-Planckian \cite{rytov,polder}. The super-Planckian RHT at some frequencies means that the spectrum of RHT between two parallel surfaces having temperatures $T_{\rm{E}}$ and $T_{\rm{C}}<T_{\rm{E}}$ exceeds the spectrum of RHT between two parallel surfaces of two black bodies with same temperatures. In the last case RHT does not depend on the distance $d$ and reproduces the Planckian radiation spectrum.

The efficient involvement of evanescent waves into RHT across a subwavelength gap is a resonant process. A pair of mutually coupled surface-phonon polaritons excited in the IR range at the interfaces of media 1 and 3 with medium 2 can dramatically enhance the RHT making it strongly super-Planckian \cite{pendry,hu,greffet,ottens}. This effect resulted in a scientific direction called near-field TPV systems (see Refs.~ \cite{Bauer,joannopoulos,ben,zhang,lee}). Near-field TPV systems cannot be physically split onto an emitter and an absorber of emission, and the RHT process is not anymore the two-stage one. Obviously, the conventional Stefan-Boltzmann and Wien laws restricting the TPV systems and resulting in the requirement of high temperatures become not applicable. In a near-field TPV system one may try to obtain a high electric output with rather low $T_{\rm{E}}$ \cite{NFTPV,NFTPV1}. Of course, the efficient PV conversion still requires the strong frequency selectivity of the RHT. A conventional filter is not helpful, because we need photon tunneling from the emitter to the PV cell which will be blocked by the filter. However, both emitter and PV cell of a near-field TPV system can be structured so that to engineer the resonant RHT in the predefined frequency range. A theoretically successful attempt in this direction is based on the application of graphene sheets \cite{ben}.
\begin{figure}[t!]
\raggedright
\subfigure[]{
\includegraphics[width=7.3cm]{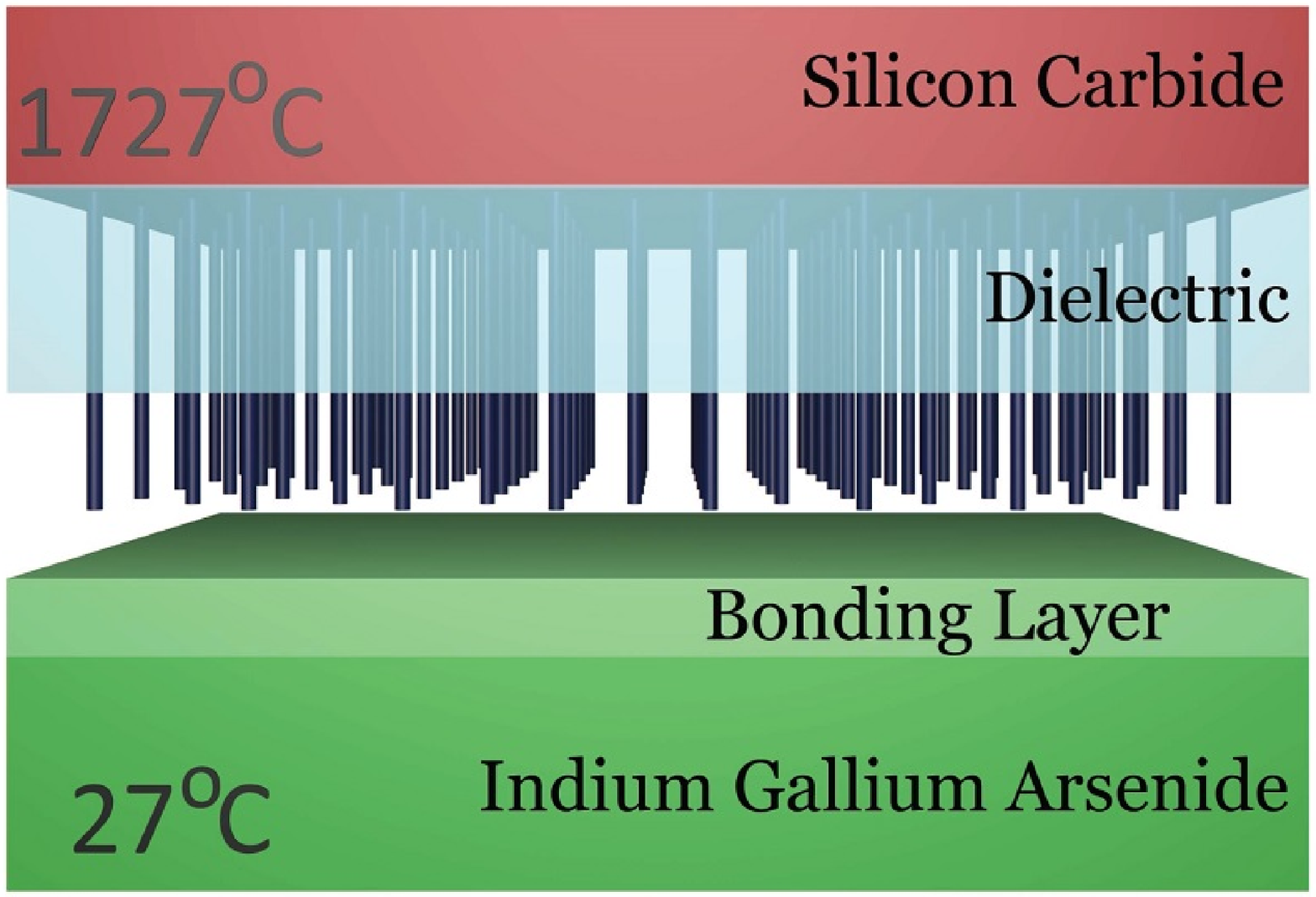}
\label{fig:P_structure}}
\subfigure[]{
\hspace{-0.2cm}
\includegraphics[width=8.5cm]{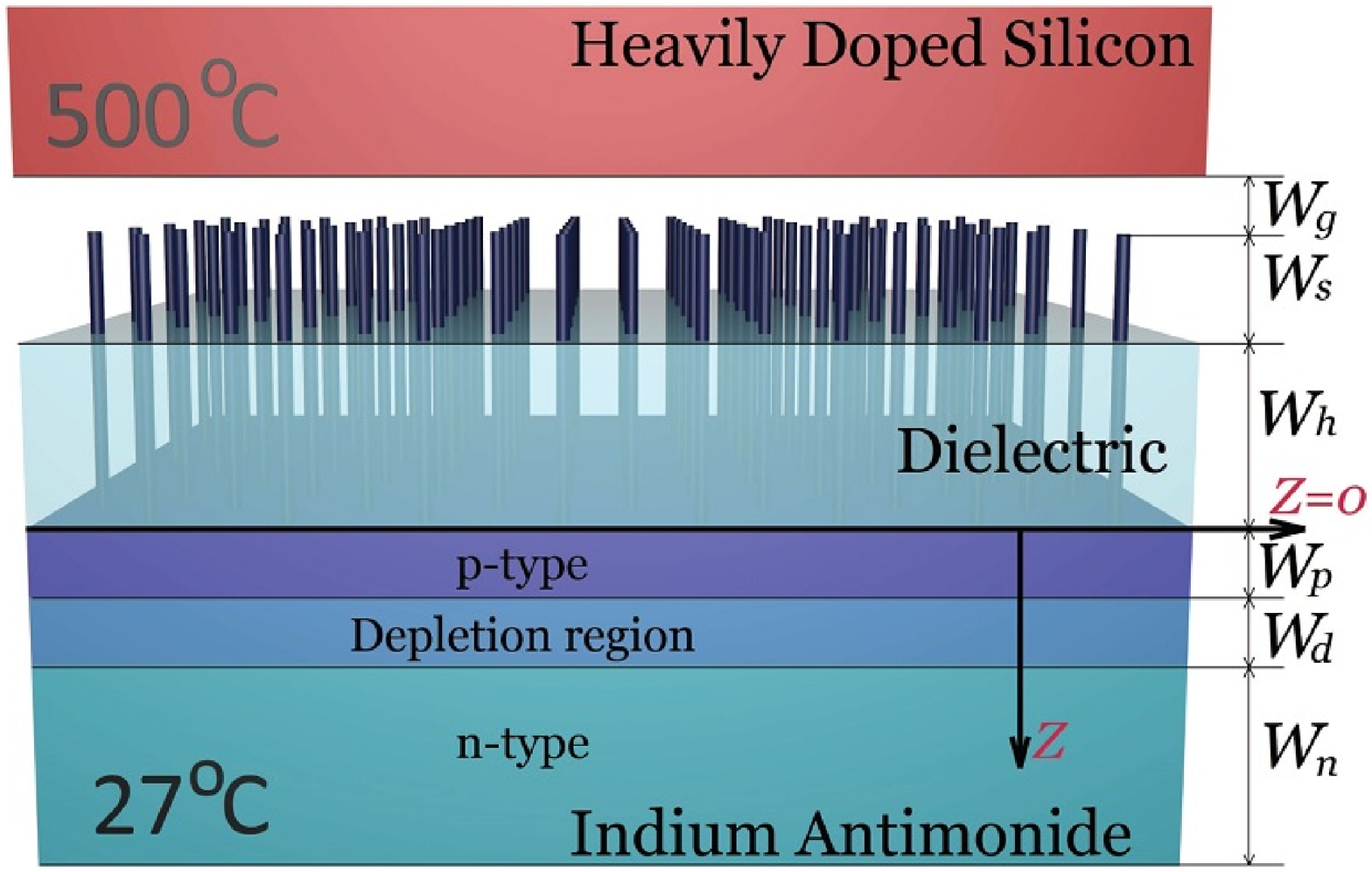}
\label{fig:structure}}
\caption{(a)--Previous structure with hot nanowires. (b)--New structure with cold nanowires.}
\end{figure}

However, near-field TPV systems requiring the gap as tiny as $d=10-20$ nm are still impractical for the electricity production. The parallelism of two interfaces so close to one another is presently feasible only in the area of several square microns \cite{NFTPV1}. With $d>20-30$ nm the RHT cannot be strongly super-Planckian \cite{NFTPV1,Bauer}. Therefore, in our group we have developed, starting from our initial work \cite{nefedov} the theory of strongly super-Planckian RHT in TPV systems with a micron or slightly submicron gap. Such systems called microgap TPV systems imply the vacuum gap $d\ge 0.5\ \mu$m.  Starting from the invention \cite{MTPV}, this technology (based on tubular quartz spacers and adjusting springs) has developed to practical TPV devices with the working area of square decimeter \cite{MTPV1,MTPV2}. In order to achieve the strongly super-Planckian RHT in such structures we suggested to grow nanowires at the sides of the micron gap \cite{nefedov}. In our further works \cite{maslovski,simovski,mirmoosa}, we have developed this approach and shown that the narrow-band strongly super-Planckian RHT is achievable in such systems with feasible design parameters. Finally, in Ref.~\cite{mirmoosa1}, we have studied the power output and optimized the whole structure imposing some practical restrictions (no energy-consuming cooling and no challenging interdigital geometry of the wire medium). In Ref.~\cite{mirmoosa1}, we obtained an unprecedent ($3.3\times10^4$ W/m$^2$) power output on the condition that the PV cell temperature is maintained by the usual tap water. This structure (the cooling frame is not shown) is schematically depicted in Fig.~\ref{fig:P_structure}. The photon tunneling occurred in our model across the tiny ($d=10$ nm) gap between the ends of hot tungsten nanowires connected to the hot silicon carbide plate. The wire medium supports propagating the huge spatial spectrum of thermal radiation. Spatial harmonics with
transversal (in-plane) wave numbers $q>k\equiv \omega/c$ (which in the vacuum gap correspond to evanescent waves) are propagating in the wire medium \cite{mario,simovski2,kivshar}. This feature of the wire medium allows a huge resonant RHT across the structure. It occurs in a predefined narrow band of frequencies. To design the structure we have used the method \cite{maslovski}, where the radiative heat is expanded onto both frequency $\omega$ and spatial $q$ spectrum, and for each spatial harmonics the impedance matrices of all effective layers of the structure are simulated. The method takes into account the generation of radiative heat in all effective layers and fits the alternative method based on the Green function of the whole structure \cite{maslovski,mirmoosa}. Though the method of the Green function \cite{green} is more popular, our method seems to be more suitable for the design of frequency-selective TPV systems. Our method delivers the surface impedances at the internal interfaces of the structure as intermediate results. This is very important for the optimization since our design strategy is to match the impedances at both sides of every interface in the predefined frequency range. If at one frequency the perfect matching is achieved, then far from this frequency (i.e. beyond the operation band of the PV cell) the spectrum of RHT turns out to be suppressed due to the strong impedance mismatch \cite{maslovski}.

\section{Our new design and numerical results}

The target of our previous work \cite{mirmoosa1} was different from the present one. We aimed to theoretically beat the record for the power output previously claimed in the literature for the same $T_{\rm E}=2000^{\circ}$K and persuade the reader that it is feasible. Therefore, the design solution shown in Fig.~\ref{fig:P_structure} does not allow us to achieve the present goal. Moreover, the lifetime of the structure \cite{mirmoosa1} would be impractically short due to the thermal emission of hot electrons from tungsten nanowires. Finally, in Ref.~\cite{mirmoosa1}, we did not analyze the photocurrent generation using, instead, an estimation technique recommended in \cite{Bauer}. However, the most important is that we suggest a new structure operating with much lower $T_{\rm E}$ with high electric output. Moreover, we present a more reliable analysis of the resonant super-Planckian TPV performance, using the minority-carrier transport model \cite{park,zhang}.

In our new TPV system the wire medium (still of tungsten nanowires) is grown on top of the semiconductor layer and, therefore, has a room temperature, e.g.~27$^{\circ}$C. Figure~\ref{fig:structure} schematically shows this structure. Instead of SiC enhanced by W nanowires as in Ref.~\cite{mirmoosa1}, we suggest a flat emitter of heavily doped Si (doping level $2\times10^{20}$ cm$^{-3}$). This geometry allows the coupled SPP to be excited at the interfaces silicon-vacuum and wire medium-vacuum. Calculations based on the effective-medium model for the wire medium of W nanowires \cite{mario} have shown that the band where the SPP is excited is adjustable via the design parameters of the wire medium. The involvement of SPP combines the advantages of our reference design Fig.~\ref{fig:P_structure} and those of the near-field TPV generator suggested in Ref.~\cite{ben}. Really, in our reference design the role of the wire medium was only to effectively approach the emitter to the PV cell. Therefore, the RHT though super-Planckian was not very strong, and the gain compared to the black-body RHT in the operation band was nearly equal 6. Now, the wire medium layer is strongly coupled to the emitter via the SPP, and the expected gain is much higher.

Optical parameters of heavily-doped silicon at $T_{\rm{E}}=500^{\circ}$C were taken from Ref.~\cite{basu}. The array of nanowires is assumed to be grown in the dielectric medium with thickness $W_{\rm{h}}$ and permittivity $\varepsilon_{\rm{h}}$ to be found in the numerical optimization. Parameters of W were taken from \cite{ordal}. An array of tungsten nanowires partially  hosted by the dielectric and partially free-standing is replaced in our calculations by two homogeneous layers of effective uniaxial media. One layer corresponds to the free-standing parts of nanoiwres, another layer -- to the hosted parts. Both these effective media have hyperbolic dispersion. Here we use the effective-medium model whose application for the analysis of RHT was validated in our previous work \cite{mirmoosa}. In principle, nanowires can be made of another materials such as gold. However, for gold nanowires the simple effective-medium model turns out to be in accurate \cite{mirmoosa}. So, the choice of tungsten allowed us to avoid full-wave simulations evaluating the impedance matrices of wire-medium layers.

The tiny gap ($W_{\rm{g}}$) prevents the harmful contact of nanowires and the emitter. The realistic fabrication tolerances of wire media \cite{simovski2} impose the restriction $W_{\rm{g}}\ge 10$ nm. The gap between the emitter and the dielectric interface $d=W_{\rm{g}}+W_{\rm{s}}$ is taken minimal possible in the microgap technology \cite{MTPV1,MTPV2} i.e. equals 500 nm. The volume fraction of nanowires ($f_{\rm{v}}=\pi r_0^2/{a^2}$, where $r_0$ and $a$ are the radius of each nanowire and the array period, respectively) was varied in our calculations from 0 to 0.3 in order to adjust the band of SPP combining in with the band of the best matching. As it was explained above, we chose InSb as a material for the PV cell. Complex permittivity of the doped InSb with n and p types of conductivity was taken from~Refs.~\cite{ben,lee,gobeli}.

The total photocurrent per unit area ($J_{\rm{ph}}$) is the sum of the drift current (of photovoltaic charge carriers) and the gradient current (due to the gradient of minority carriers concentration) \cite{park,zhang}:
\begin{equation}
J_{\rm{ph}}=J_{\rm{d}}+J_{\rm{n,p}}=\displaystyle\int^\infty_{\omega_{\rm{g}}}\displaystyle\left(j_{\rm{d}}(\omega)+j_{\rm{n,p}}(\omega)\right)d\omega.
\label{eq:photocurrent}
\end{equation}
In the above equation we have \cite{park,zhang}:
\begin{equation}
\begin{split}
&j_{\rm{n,p}}(\omega)=e_0\left[D_{\rm{e}}\vert\displaystyle\frac{dn_{\rm{e}}(\omega,z=z_{\rm{p}})}{dz}\vert+D_{\rm{h}}\vert\displaystyle\frac{dn_{\rm{h}}(\omega,z=z_{\rm{n}})}{dz}\vert\right],\\
&j_{\rm{d}}(\omega)=\displaystyle\frac{e_0}{2\pi\hbar\omega} \int_0^{\infty} Q(\omega,q)\left[e^{-2{\rm{Im}}(\beta)z_{\rm{p}}}-e^{-2{\rm{Im}}(\beta)z_{\rm{n}}}\right] qdq,
\end{split}
\label{eq:current_density}
\end{equation}
where $e_0$ is the electron charge, $D_{\rm{e,h}}$ is the diffusion constant, $z_{\rm{n,p}}$ are coordinates of the edges of the depletion region, and ${\rm{Im}}(\beta)$ is the imaginary part of the vertical component of the wave vector in InSb: $\beta=\sqrt{k_{\rm InSb}^2-q^2}$, where $q$ is the transverse component of the wave vector of spatial harmonic (see above), and $k_{\rm InSb}$
is the complex wave number of InSb at frequency $\omega$. In Eq.~\ref{eq:current_density}, $Q(\omega,q)$ is the frequency-and-spatial spectrum of RHT i.e. power absorbed by the unit area of the PV cell per unit interval of frequencies and unit interval of $q$. Finally, $n_{\rm{e,h}}$ denote the minority carrier concentrations. These concentrations are solutions of the equation \cite{park}:
\begin{equation}
D_{\rm{e,h}}\displaystyle\frac{d^2n_{\rm{e,h}}}{dz^2}-\displaystyle\frac{n_{\rm{e,h}}}{\tau_{\rm{e,h}}}+{1\over \pi}\int_0^\infty {\rm{Im}}(\beta)Q(\omega,q)e^{-2{\rm{Im}}(\beta)z}qdq=0.
\label{eq:steady}
\end{equation}
Here, $\tau_{\rm{e,h}}$ is the relaxation time. To calculate $Q(\omega,q)$ we used the method~\cite{maslovski} as we did in \cite{mirmoosa1}.
The solutions of Eqs.~\ref{eq:steady} were obtained imposing four  boundary conditions: two for $n_e$ (at $z=0$ and at $z=W_{\rm p}$) and two for holes (at $z=W_{\rm p}+W_{\rm d}$ and at $z=W_{\rm p}+W_{\rm d}+W_{\rm n}$). The first-type condition is $D_{\rm{e,h}}(dn_{\rm{e,h}}/dz)=S_{\rm{e,h}}n_{\rm{e,h}}$ (at $z=0$ for electrons and at $z=W_{\rm{p}}+W_{\rm{d}}+W_{\rm{n}}$ for holes), where $S_{\rm{e,h}}$ is the surface recombination rate.
The second-type condition is $n_{\rm{e,h}}=0$ at $z=W_{\rm p}$ for electrons and at $z=W_{\rm p}+W_{\rm d}$ for holes. Solving Eq.~\ref{eq:steady} we find both components of the photocurrent in  Eq.~\ref{eq:photocurrent}. In accordance to the equivalent circuit in Fig.~\ref{fig:TPV}, we find the total current collected from unit area $J$ (current in the load divided by the illuminated area of the PV cell \cite{Bauer}) as the difference $J=J_{\rm{diode}}-J_{\rm{ph}}$, in other words,
\begin{equation}
\begin{split}
J=\left[e_0n_{\rm{i}}^2\left(\displaystyle\frac{D_{\rm{e}}}{N_{\rm{p}}L_{\rm{e}}}+\displaystyle\frac{D_{\rm{h}}}{N_{\rm{n}}L_{\rm{h}}}\right)\right]\left[\exp\left(\displaystyle\frac{eV}{K_{\rm{B}}T_{\rm{C}}}\right)-1\right]-J_{\rm{ph}},
\end{split}
\label{eq:Imax}
\end{equation}
in which $K_{\rm{B}}$ is the Boltzmann constant, $n_{\rm{i}}$ denotes the intrinsic carrier concentration, $N_{\rm{n,p}}$ represents the carrier concentration, $L_{\rm{e,h}}=\sqrt{D_{\rm{e,h}}\tau_{\rm{e,h}}}$ is the diffusion length and $V$ is the load voltage which varies from zero (short-circuit case) to its maximum value as the load is considered to be infinity (open-circuit case). Table~\ref{tab:properties} contains the parameters that we use to calculate $J$ (data are taken from works \cite{gonzales_cuevas, francoeur, chen}).
\begin{table}[t!]
\caption{Input parameters for calculating the photocurrent density and the current density referred to the load}
\raggedright
	\begin{tabular}{p{5cm} p{3cm}}
		\hline\noalign{\smallskip}
                      p-region & n-region\\[1ex]
                      \hline\noalign{\smallskip}
                      $D_{\rm{e}}=186\,{\rm{cm}^2}{\rm{s}^{-1}}$ & $D_{\rm{h}}=5.21\,{\rm{cm}^2}{\rm{s}^{-1}}$\\
                      $\tau_{\rm{e}}=1.45\,$ns & $\tau_{\rm{h}}=1.81\,$ns\\
                      $S_{\rm{e}}=10^4\,{\rm{m}}{\rm{s^{-1}}}$ & $S_{\rm{h}}=0\,{\rm{m}}{\rm{s}^{-1}}$\\
                      $W_{\rm{p}}=100\,$nm & $W_{\rm{n}}=10\,\mu$m\\
                      $N_{\rm{p}}=10^{19}\,{\rm{cm}}^{-3}$ & $N_{\rm{n}}=10^{19}\,{\rm{cm}}^{-3}$\\ [1ex]
		\hline\noalign{\smallskip}
$n_{\rm{i}}=2\times10^{16}{\rm{cm}}^{-3}$ and $W_{\rm{d}}=10\,$nm\\
\hline
\end{tabular}
\label{tab:properties}
\end{table}
For the optimal load, the electric power density ${\cal P}_{\rm{elec}}=J\cdot V$ gets the maximum value. Then we can obtain the photovoltaic conversion efficiency as the ratio between the maximal (over all possible loads) electric power output per unit area of the cell to the integral value of radiative heat transfer ${\cal P}_{\rm radiative}$ -- power flux at the interface $z=0$ between the dielectric layer and InSb. In other words, we have:
\begin{equation}
\eta=\displaystyle\frac{{\rm{Max}}\left[{\cal P}_{\rm{elec}}\right]}{{\cal P}_{\rm{radiative}}}=\displaystyle\frac{{\rm{Max}}\left[J\cdot V\right]}{\displaystyle\int_0^\infty{d\omega}\displaystyle\int_0^\infty{Q(\omega,q)qdq}}.
\label{eq:efficiency}
\end{equation}
\begin{table}[t!]
\caption{Optimized values for the parameters of the structure under study}
\raggedright
	\begin{tabular}{p{5.1cm}p{3cm}}
		\hline\noalign{\smallskip}
                      Doping concentration of silicon & $2\times10^{20}\,{\rm{cm^{-3}}}$\\
                      Radius of nanowires & $30\,$nm\\
                      Period of nanowires array & $100\,$nm\\
                      Relative dielectric constant of host medium for nanowires & $2.25$\\
                      $W_{\rm{g}}$ & $10\,$nm\\
                      $W_{\rm{s}}$ & $500\,$nm\\
                      $W_{\rm{h}}$ & $730\,$nm\\ [1ex]
		\hline
\end{tabular}
\label{tab:values}
\end{table}
\begin{figure}[t!]
\hspace*{0cm}
\subfigure[]{
\includegraphics[width=7.5cm]{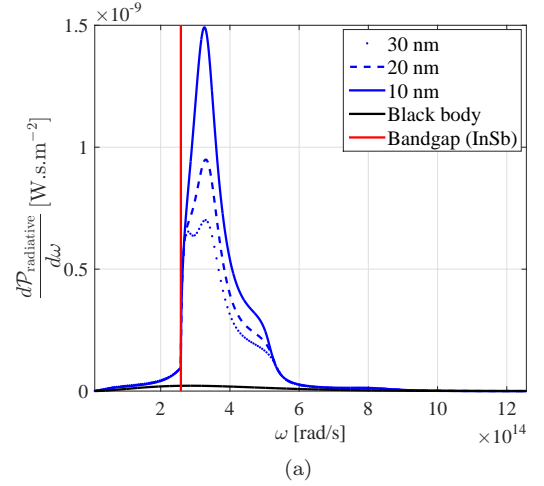}
\label{fig:power_density}}
\subfigure[]{
\includegraphics[width=7.5cm]{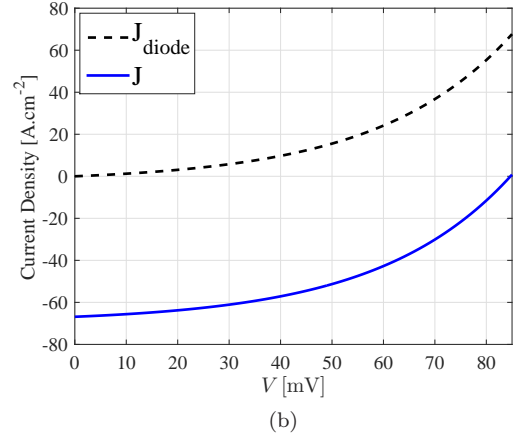}
\label{fig:iv}}
\caption{(a)--Spectrum of radiative heat transfer at $z=0$. The blue and black curves correspond to our optimized structure and to the black-body RHT, respectively. The dashed and dotted curves show the effect of the thickness of the tiny gap $W_{\rm{g}}$. (b)--The current density over the load voltage ($J$--$V$ curve) (here, $W_{\rm{g}}=10$ nm).}
\end{figure}
Table~\ref{tab:values} illustrates the optimized values for the structure parameters. The spectrum of radiative heat transfer (power absorbed per unit area of the PV cell per unit interval of angular frequencies) is shown in Fig.~\ref{fig:power_density}. This spectrum is shown in comparison with that of the black-body RHT. We see that it is much more narrow-band than the black-body RHT and by an order of magnitude exceeds it in the band between $\omega_{\rm{min}}=2.6\times10^{14}$ rad/s and $\omega_{\rm{max}}=5\times10^{14}$ (the band-gap frequency of the PV material is marked by red line). The spectral maximum at which the RHT spectrum exceeds the black-body one by two orders of magnitude holds at $\omega_{\rm{M}}\approx 3.3\times10^{14}$ rad/s, and the crucial part of the resonant spectrum is above $\omega_{\rm g}$. As Fig.~\ref{fig:power_density} indicates, increasing $W_{\rm{g}}$ from 10 nm to 20- and 30 nm causes the decreasing in the maximum value at $\omega_{\rm{M}}$. 

For comparison we have calculated RHT in our structure removing the nanowires. The result is visually very close to the black-body RHT. Therefore, the difference between two curves in Fig.~\ref{fig:power_density} is practically equal to the gain granted by nanowires. Figure~\ref{fig:iv} shows the corresponding Volt-Ampere characteristics of our PV cell ($J$--$V$ curve). The black and blue curves are diode current and the total current, respectively. At the optimal load, the electric power is equal ${\cal P_{\rm{elec}}}=26\,$kW/m$^{2}$ corresponding to the total (photovoltaic) conversion efficiency $\eta=13${\%}. Work~\cite{ben} has reported the PV conversion efficiency 24{\%} for a near-field TPV system also operating at 500$^{\circ}$C. In that structure, the gap between the emitter (hexagonal boron nitride) and the semiconductor (indium antimonide) was equal 16 nm (also an atomic layer of graphene with the chemical potential of 0.5 eV was located on the surface of InSb). Clearly, this structure essentially refers to near-field TPV systems, whereas we claim the results of the same order for a microgap TPV system.

\begin{table}[t!]
\caption{Overall conversion efficiency versus $W_{\rm h}$.
}
\raggedright
	\begin{tabular}{p{1.1cm}p{0.8cm}p{0.8cm}p{0.8cm}p{0.8cm}p{0.8cm}p{0.8cm}p{0.8cm}p{0.8cm}}
		\hline\noalign{\smallskip}
                      $W_{\rm{h}}\,$[nm] & 300 & 400 & 500 & 600 & 700 & 800 & 900 & 1000\\[1ex]
                      \hline\noalign{\smallskip}
                      $\eta{\%}$ & 16.3 & 15.7 & 14.8 & 13.9 & 13.2 & 12.7 & 12.4 & 12.1\\[1ex]
\hline\noalign{\smallskip}
\end{tabular}
\label{tab:length}
\end{table}

Moreover, our results can be improved if we soften the restriction for the thickness of the host dielectric. We have taken above $W_{\rm{h}}=730\,$nm that is approximately 60{\%} of the total length $W_{\rm{h}}+W_{\rm{s}}$ of a nanowire because such free-standing wire media were reported in the literature (see e.g.~Ref.~\cite{simovski2}). However, there are no physical reasons prohibiting smaller values for the hosted length $W_{\rm{h}}$ than 60\% of the total length for a so robust metal as tungsten.  As Table~\ref{tab:values} indicates, decreasing the thickness $W_{\rm{h}}$ from 730 nm to 300 nm results in the increasing of the conversion efficiency from 13{\%} to 16.30{\%}. Nanowires with a free-standing portion longer than the hosted portion are not yet reported in the literature, however, we believe that they are feasible. In the case of $W_{\rm{h}}=300\,$nm, the maximum electric power is about 40.64 kW/m$^{2}$. By fixing $W_{\rm{h}}=300\,$nm and trying to change other parameters of the structure, we can even match better the emitter to the semiconductor near its band gap. The best case is when the parameters in Table~\ref{tab:values} remain the same except the relative dielectric constant of the host medium. If this constant changes from 2.25 to unity (a transparent dielectric at infrared), the conversion efficiency and the maximum output power increase to 18.7{\%} and 55.2 kW/m$^{2}$. These values are closer to those claimed in Ref.~\cite{ben}.

Finally, in comparison with the practical macroscopic system, where the heat-to-electricity conversion resulted in the power output 5.5 W/m$^{2}$ for the temperature 500$^{\circ}$C \cite{TE}, our 55.2 kW/m$^{2}$ mean the gain $10^4$. This amazing gain, to our opinion, will justify the fabrication costs of a nanostructured micro-gap TPV system.

\medskip
\section{Conclusions}

In the present work we have introduced a new type of efficient and feasible low-temperature TPV system with very high electric output. The system is based on the existing microgap technology 
for TPV systems, and, therefore, potentially allows one to collect the photocurrent from macroscopic areas covering the emitter with temperatures $400-600^{\circ}$C. The use of the realistic tungsten wire medium grown on top of an IR PV cell together with the heavily-doped silicon emitter results in our calculations in the excitation of a surface-phonon-polariton at the effective internal interfaces of the system. Then the wire medium not only effectively approaches the emitter and the TPV cell to one another, but also couples them in a resonant way. The frequency selectivity is achieved not only due to the effectively multi-layer structure of the TPV system and high contrast between layers as in our precedent works. It is enhanced by the resonant excitation of the SPP, that results in further frequency squeeze of the spectrum of radiative heat transfer and enhancement of its maximum. A huge gain -- four orders of magnitude -- for the electric power output compared to commercial analogues operating at same temperatures is obtained. The next stage of our studies will be the further decrease of $T_{\rm E}$ which will allow the TPV electric generators to replace the thermoelectric ones in the electric power supplies, computer processors, etc.

\end{document}